\documentstyle[12pt]{article}

\makeatletter
\@addtoreset{equation}{section}
\makeatother

\def\RR{{{\rm l}\kern-.17em{\rm R}}}
\newcommand\bfab{\em}
\newcommand\Gh{{\hat{G}}}

\newcommand\xh{{\hat{x}}}

\newcommand\sbr{\overline{s}}
\newcommand\wbr{{\overline{w}}}
\newcommand\Dub{{\overline{\Delta}_N}}
\newcommand\Dlb{{\underline{\Delta}_N}}

\newtheorem{thm}{Theorem}[section]
\newtheorem{lemma}[thm]{Lemma} 
\newtheorem{cor}[thm]{Corollary} 
\newcommand{\EQ}{\begin{equation}}
\newcommand{\EN}{\end{equation}}
\newcommand{\BEQ}{\begin{equation}}
\newcommand{\NEQ}{\end{equation}}
\newcommand{\nn}{\nonumber}

\def\ni{\noindent}

\def\gapprox{\stackrel{>}{\sim}\,}
\def\eopp{{$\Box$}}      
\def\xbf{{\bf x}\,}
\def\Ebf{{\bf E}\,}

\def\Qt{\tilde{Q}\,}
\def\wt{\tilde{w}\,}
\def\fl{{f^{(\ell)}}}
\def\flp{{f^{(\ell+1)}}}
\def\fh{\hat{f}}
\def\fhj{\hat{f}^{(j)}}
\def\fhl{\hat{f}^{(\ell)}}
\def\fhlp{\hat{f}^{(\ell+1)}}
\def\fhjN{\widehat{f}_N^{(j)}}
\def\fhlN{\hat{f}_N^{(\ell)}}
\def\fhlpN{\hat{f}_N^{(\ell+1)}}
\def\eps{{\epsilon}}

\def\part{\partial}

\def\Kh{\hat{K}}
\date{ }

\begin{document}
\bibliographystyle{plain}

\title{Piecewise Convex Function Estimation: \\ 
Pilot Estimators} 

\author{Kurt S.~Riedel \\
Courant Institute of Mathematical Sciences \\
New York University}

\maketitle
\begin{abstract} 
Given noisy data, function estimation is considered when the unknown
function is known {\it a priori} to consist of a small number of regions
where the function is either convex or concave. 
When the number of regions is unknown, the
model selection problem is to determine the number of convexity change
points. For kernel 
estimates in Gaussian noise, the number of 
false change points is evaluated as a function of the smoothing
parameter.  
To insure that the number of false convexity change points tends to zero,
the smoothing level must be larger than is generically optimal for minimizing
the mean integrated square error (MISE).
A two-stage estimator is proposed and shown to achieve
the optimal rate of convergence of the MISE.    
In the first-stage, the number and location of the change points
is estimated using strong smoothing. In the second-stage, a 
constrained smoothing
spline fit is performed with the smoothing level  chosen to minimize the
MISE. 
The imposed constraint is that a single  change point
occur in a region about each empirical change point from the 
first-stage estimate. This constraint is equivalent to the requirement
that the third derivative of the second-stage estimate have a
single sign in a small neighborhood about each first-stage
change point.
The change points from the second-stage are 
in a  neighborhood of the first-stage change points,
but need not be at the identical locations. 
\end{abstract}


\section{Introduction}\label{I}

Our basic tenet is:
``Most real world functions are piecewise $\ell$-convex with a small number of
change points of convexity.''
Given $N$ measurements of the unknown function, $f(t)$, contaminated with
random noise, we seek to estimate $f(t)$ while preserving the geometric
fidelity of the estimate, $\fh(t)$, with respect to the true function. In
other words, the number and location of the change points of convexity of
$\hat{f}(t)$ should approximate those of ${f}(t)$.
We say
that $f(t) $ has $K$ change points of $\ell$-convexity  with change points
$x_1 \leq x_2 \ldots \leq x_K$ if $(-1)^{k-1} f^{(\ell )} (t) \geq 0$ for
$x_k \leq t \leq x_{k+1}$.  

The idea of constraining the function fit to preserve {\em prescribed}
$\ell$-convexity
properties has been considered by a number of authors
\cite{Maechler,MU85,Utreras,VW}. 
The more
difficult problems of determining the number and location of the
$\ell$-convexity breakpoints will be a focus of this article. 
Historical perspectives to the problem may be found in \cite{Li95,Mammen91}.
``Bump hunting'' dates back at least to \cite{GG}. 
Silverman \cite{Silverman83}
and Mammen et al.\ \cite{Mammen,MMF}
formulated the problem as a sequential hypothesis testing problem.
We refer to the estimation of the number of change points as the
``model selection problem'' because it resembles model selection in an
infinite family of parametric models.

An interesting result of \cite{Mammen,MMF,RiedLet} is that kernel smoothers
will often produce too many inflection points/wiggles. If the amount of
smoothing is chosen to minimize the mean integrated square error (MISE),
then with nonvanishing asymptotic probability, the estimate will
have multiple inflection points in a neighborhood of an actual one.
For many applications, estimating the correct shape is more
important than minimizing the MISE.

In this article, we propose a  class of two-stage estimators which
estimate the $\ell$-change points in the first-stage and then
perform a constrained regression in the second-stage. In the first
stage, the function is strongly smoothed while the  smoothing in
the constrained second-stage is optimized for the minimal mean square
error.
When the change points are correctly specified, the constrained spline
estimate has a smaller square error (as measured in a particular norm)
than the unconstrained estimate. 


Our second-stage estimate achieves the asymptotically 
optimal MISE convergence rate
while suppressing artificial change points that can occur with the
unconstrained method. Our proof does not exclude the possibility that the 
second-stage estimate has spurious inflection points far from the
first-stage inflection points.
We believe that our estimator
has the same {\em relative} convergence rate 
as standard nonparametric methods.
Thus our estimator suppresses artificial wiggles at nontrivial 
computational costs but no lost of MISE. 
 
In Section \ref{MISE}, 
we show that the constrained smoothing
spline estimate achieves the optimal rate of convergence for the expected
square error even when the constraints are occasionally misspecified,
provided that the misspecification rate is sufficiently small.

In Section \ref{WLEsect}, we evaluate the expected number of 
false (extra) empirical change points \cite{MMF}
for kernel smoothers 
when the errors are Gaussian.
By adjusting the smoothing parameter, we can guarantee an asymptotically
small probability of an error in our imposed constraints.

In Section \ref{Dadapt},
we propose two-stage estimators which estimate the number and location
of the $\ell$-change points in the first-stage. 
In the second-stage,  
we impose that $\hat{f}^{(\ell+1)}$ be 
positive/negative in a small region about each of the empirical 
$\ell$-change points. The main advantage of the two-stage procedure
is that less smoothing is required in the second-stage than in the
first-stage while preserving the geometric fidelity.

Section \ref{Pot Ext} discusses potential extensions of the
method.
Section \ref{PCICSec} describes a global shape optimization
that is \emph{heuristically} designed to be efficient in the amount of 
smoothing subject to determining the number of change points consistently.

\section{Expected error under inexact convexity constraints} \label{MISE}

We are given $N$ measurements of the unknown function, $f(t)$:
\begin{equation}\label{M1}
 \ y_i = f(t_i)  + \epsilon_i \ .
\end{equation}
The mean integrated squared error (MISE) for the estimate, $\hat{f}^{(j)}$, 
of the $j$th derivative  is defined as
${\bf E} \left[ \int |\hat{f}^{(j)}_{N,\lambda}(t)-f^{(j)} (t)|^2 \right]$.
We consider the MISE for constrained estimation as the number
of measurements, $N$, tends to infinity.
In describing the large $N$ asymptotics, 
we consider a sequence of measurement problems. For each $N$,
the measurements occur at $\{ t_i^N, \ i= 1 \ldots N \}$,
We suppress the superscript, $N$,
on the measurement locations $t_i \equiv t_i^N$. 
We define the empirical
distribution of measurements, $F_N(t)= \sum_{t_i \le t} 1/N$, 
and let $F(t)$ be its limiting distribution.

\noindent
{\bf Assumption A} 
{\em Consider the sequence of
estimation problems: $y_i^N = f(t_i^N) + \eps_i^N$,
where the $\eps_i^N$ are zero mean random variables and 
${\bf Cov}[\eps_i^N,\eps_j^N] =\sigma^2 \delta_{i,j}$.
Assume that the distribution of  measurement locations, 
converges in the sup norm: $|F_N(t)- F(t)| \rightarrow 0$,
where
$F(t)$, is $C^{\infty}[0,1]$ and $0<c_F<F'(t)<C_F$. 
}

We measure the convergence of a set of measurement times to
the continuum limit in terms of the discrepancy of the point set:

\noindent
{\bf Definition} {\em The star discrepancy of $\{ t_1 \ldots t_N \}$ 
with respect to the continuous distribution $F(t)$ is
$D^*_N \equiv\ \sup_t \{ F_N(t) -F(t) \}$.
}

Equivalently, $D_N^* \sim 1/2N + \max_{1\le i\le N} | F(t_i) - (i-\ 1/2)/N |$.
For regularly spaced points, 
$D_N^* \sim 1/N$, while for randomly spaced points, 
$D_N^* \sim \sqrt{\ln[\ln[N]]/N}$ by the Glivenko-Cantelli Theorem.

A popular linear estimator is the smoothing spline \cite{Wahba91}:
$\fh = \arg\min {\rm VP}[f ] $, where
\begin{equation}\label{3.21}
{\rm VP}[f ] \equiv
\frac{\lambda}{2}\int |f^{(m)} (s) |^2 ds +
\frac{1}{N} 
\sum_{i=1}^N \frac{|y_i- f(t_i)|^2}{\sigma^2} 
\ . \ \
\end{equation}
We denote the standard Sobolev space of functions with square integrable
derivatives by $W_{m,2}$ \cite{Wahba91}. 
In general, the smoothing parameter will be decreased as $N$ increases:
$\lambda_N \rightarrow 0$. 
For smoothing splines, we add the stronger requirements:

\noindent
{\bf Assumption A$^*$ (Cox \cite{Cox})} {\em 
Let Assumption A hold with $f\in W_{m,2}$ and $m > 3/2$. 
Consider the sequence of 
smoothing spline minimizers of  (\ref{3.21}).
Let the smoothing parameter, $\lambda_N$, satisfy $\lambda_N \rightarrow 0$
and $D_N^* \lambda_N^{-5/(4m)} \rightarrow 0$  
as $N\rightarrow \infty$.}

The constraint that $D_N^* \lambda_N^{-5/(4m)} \rightarrow 0$  
is very weak since the optimal value of $\lambda_N$ satisfies
$\lambda_N \sim N^{-2m\/ (2m+1)}$.
The assumption that $F(t)$ is $C^{\infty}[0,1]$ can be weakened.

For unconstrained  smoothing spline estimates, the 
MISE has the following upper bound:

\begin{thm}[Ragozin \cite{Ragozin}, Cox \cite{Cox}] \label{Ragthm}
Let Assumption A$^*$ hold, 
and denote  the smoothing spline estimate from (\ref{3.21})
by  $\hat{f}_{N,\lambda} (t)$. 
As $N\rightarrow\infty$,
\begin{equation}\label{6.1}
{\bf E} \left[ 
\int |\hat{f}^{(j)}_{N,\lambda} (t)-f^{(j)} (t)|^2 \right] \le
\alpha_j \lambda^{{(m-j)\over m} } \parallel f \parallel_m^2 
+ {\beta_j\sigma^2 \over
N\lambda^{{(2j+1)\over 2m}} } \ \ ,
\ j \le m \ ,
\end{equation}
where $\alpha_j$ and $\beta_j$ are positive constants.  The MISE  is
minimized by $\lambda = O(N^{-2m/ (2m+1)})$ and
\begin{equation}\label{6.1b}
E \left[ \int |\hat{f}_{N,\lambda}^{(j)} (t)-f^{(j)} (t)|^2 dt \right] =
O\left( N^{{-2(m-j ) \over (2m+1)}} \right) \ .
\end{equation}
\end{thm}

For  uniformly spaced measurement points, this result is in \cite{Ragozin},
while Cox \cite{Cox} generalizes the result to these more general 
conditions on the measurement points. 
Cox imposes the condition that $j\ <\ m$ while his proof applies to $j=m$ 
as well.

Given change points, $\{ x_1 ,x_2 \ldots x_K \}$, we define the closed
convex cone:
\begin{equation}\label{3.1i}
V^{K,\ell }_{m,2} [x_1 ,\ldots ,x_K ] = \{ f \in
W_{m,p}  \ | \ (-1)^{k-1}
f^{(\ell )} (t) \geq 0 \ \ {\rm for } \ \ x_{k-1} \leq t \leq x_k \} \ ,
\end{equation}
where $x_0 \equiv 0$ and $x_{K+1} \equiv 1$.
When the change points are unknown, we need to consider
$V^{K,\ell }_{m,2} = \bigcup_{\xbf \in [0,1]^K} V^{K,\ell }_{m,2} [\xbf]
\cup (- V^{K,\ell }_{m,2} [\xbf])$,
where ${\bf x} \equiv ( x_1 ,x_2 \ldots x_K )$.

If we know the change point locations, $\xbf$, a natural estimator
is $\fh = \arg\min {\rm VP}[f ] $ subject to the convex constraints
that  $\fh \in V^{K,\ell }_{m,2} [\bf{x}]$.
This constrained spline estimate is the basis of our analysis.
Detailed representation and duality results are in \cite{Riedel1}.
If we correctly impose the constraint that $f \in V_{m,2}^{K,\ell}[\bf{x}]$,
the constrained spline fit always outperforms the nonconstrained fit
as the following theorem indicates:

\begin{thm}[Utreras \cite{Utreras}] \label{Uthm}
{Let $f$ be in a closed convex cone,  $C \subseteq W_{m,2}$.
Let $\fh_u$ be the unconstrained minimizer of (\ref{3.21}) given $y_i$;
and $\fh_c$ be the constrained minimizer, 
then $\|f - \fh_c \|_V  \le \|f - \fh_u \|_V$,
where 
\begin{equation}\label{4.1}
\|f  \|_V^2 \equiv
\frac{\lambda}{2}\int |f^{(m)} (s) |^2 ds + 
{1\over N\sigma^2} \sum_{i=1}^N | f(t_i) |^2  \ .
\end{equation}
}
\end{thm}

Theorem \ref{Uthm} 
applies to any set of $y_i$ and does not use $y_i = f(t_i) +\eps_i$.
Theorem \ref{Uthm}  shows that if one is certain that $f$ is in a particular 
closed convex cone, the constrained estimate is always better than 
the unconstrained one. Unfortunately, Theorem \ref{Uthm}  does not generalize
to unions of convex cones, and thus does not apply to $V^{K,\ell}_{m,2}$.

\begin{thm}
\label{Uterthm}
Consider $f \in W_{m,2}$ 
and let Assumption A$^*$ hold. 
Consider a sequence of estimates, $\fh_{u,N}$, which satisfy the
error bound given by (\ref{6.1}), and a second sequence of estimates,
$\fh_{c,N}$, which satisfy the error bound:
$\|f - \fh_c \|_V  \le \|f - \fh_u \|_V.$
The asymptotic error bound  given by Eq.~(\ref{6.1}) holds for 
$\fh_{c,N}$ with different constants, $\alpha_j'$ and $\beta'_j$. 
\end{thm}

\noindent 
{\bfab Proof.}
For uniform sampling, this theorem is proved in \cite{Utreras} using
interpolation inequalities. To  generalize Utreras's result  to
our sampling hypotheses, we replace his Lemma 4.3 with 
(\ref{Kint1}) and  (\ref{Kint2}) (with $\delta = .05$)
and substitute $D_N^{*^{.45}}$ everywhere $1/n$ appears.
\eopp

This generalization of Utreras's result does not require the ratio of 
$\Dub \equiv \sup_{i<N} (t_{i+1} -t_{i})$ to
$\Dlb \equiv \inf_{i<N} (t_{i+1} -t_{i})$ to be bounded.
In practice, we choose our constraints empirically and sometimes impose
an incorrect constraint.
We now show that occasionally imposing the wrong constraint does not
degrade the asymptotic rate of convergence  provided that
the probability of an incorrect constraint is sufficiently small.
The following theorem is a basis for our data adaptive estimators in
Section \ref{Dadapt} .

\begin{thm}[Occasional Misspecification]\label{MSthm} 
{Consider a two-stage estimator 
that with probability, $1 - {\cal O}(p_N)$, correctly
chooses a closed convex cone $C$, with $f\in C$,
in the first-stage and then performs
a constrained regression as in 
(\ref{3.21}).
Under Assumption A$^*$, 
the  estimate, $\fhj$, satisfies the asymptotic bound (\ref{6.1})
(with different constants, $\alpha_j'$ and $\beta'_j$)
provided that as $N\rightarrow \infty$, $p_N$ vanishes rapidly enough: 
$p_N/ \lambda_N \rightarrow 0$
and $p_N N\lambda^{1\over 2m} 
\rightarrow 0$. 
}
\end{thm}
  

The proof is given in Appendix 
\ref{PROOF}.

\section{Convergence of Kernel Smoothers}
\label{WLEsect}
In this section, we examine the expected number of zeros
of a kernel smoother as a function of the halfwidth parameter.
The results in this section are proven in \cite{RiedLet}.
We begin by presenting convergence results for kernel estimators, $\fhlN(t)$,
of  $f^{(\ell)}(t)$ as $N\rightarrow \infty$. 
We define 
$\sigma^2_{N}(t) = {\bf Var}[\fhlN(t)]$,
$\xi^2_{N}(t) = {\bf Var}[\fhlpN(t)]$,
$\mu^2_{N}(t) = {\bf Corr}[\fhlN(t),\fhlpN(t)]$.
We use the notation ${\cal O_R}(\cdot)$ to denote a  size of ${\cal O}(\cdot)$
relative to the main term: 
${\cal O_R}(\cdot)=\times \ [1+ {\cal O}(\cdot)]$.
We define $\|f\|_{bv}$ to be the sum of the $L_{\infty}$ and total
variation norms of $f$.

\begin{lemma}[Generalized Gasser-M\"uller \cite{GM84,RiedLet}]\label{KS}
Let $f(t) \in C^{\ell+1}[0,1]\cap  TV[-1,1]$ 
and consider a sequence of estimation
problems satisfying Assumption A. Let
$\fhlN(t)$ be a kernel smoother estimate of the form: 
\begin{equation}\label{K1}
\fhl(t) = \frac{1}{Nh_N^{\ell+1}}
\sum_i^N \frac{y_iw_i}{F'(t_i)} \kappa^{(\ell)}({t-t_i \over h_N}) \ ,
\end{equation}
where $h_N$ is the kernel halfwidth and the weights, $w_i$
satisfy $|w_i-1| \sim {\cal O}(D_N^*/h_N)$. 
Let the kernel, $\kappa^{(\ell+1)} \in TV[-1,1]\cap C[-1,1]$,
satisfy the moment condition:
$\int_{-1}^1  \kappa(s)ds = 1$,  
and the boundary conditions:
$\kappa^{(j)}(-1) =\kappa^{(j)}(1) = 0$ for $0\le j \le \ell$.
Choose the kernel halfwidths such that $h_N\rightarrow 0$, 
and $D_N^*/h_N^{\ell +2} \rightarrow 0$; 
then $\Ebf[\fhlN](t) \rightarrow f^{(\ell)}(t)
 + {\cal O_R}(h_N + D_N^*/h_N^{\ell +1})$,
$\Ebf[\fhlpN](t) = 
\int_{-1}^1 f^{(\ell)}(t+hs)\kappa(-s)ds $  
$+\ {\cal O}(\|f\kappa^{(\ell+1)}\|_{bv} D_N^*/h_N^{\ell +2})$,
$\sigma_N^2(t)  \rightarrow \sigma^2\|\kappa^{(\ell)} \|^2 /$ 
$(N F'(t) h_N^{2\ell +1} ) +\ 
{\cal O_R}(h_N + D^*_N/h_N)$, \ 
$\xi_N^2(s) \rightarrow  \sigma^2$
$\|\kappa^{(\ell+1)} \|^2_{}$ $ / (NF'(s)h_N^{2\ell +3} ) 
+\ {\cal O_R}( h_N+ D^*_N/h_N)$, and  \ 
$\mu^2_N(t) \rightarrow {\cal O}(h_N +\  D_N^*/h_N)$ 
uniformly 
in the interval, $[h_N,1-h_N]$. 
\end{lemma}

Lemma \ref{KS} applies to all of the common kernel
smoother weightings \cite{GM84} 
such as  Priestley-Chao and Gasser-M\"uller.
This result is   slightly
stronger than previous theorems on kernel smoothers \cite{GM84}.
Our hypotheses are stated in terms of the star discrepancy while 
previous convergence theorems  \cite{GM84} 
place restrictions on both
$\Dub \equiv\ sup_{i<N}\{t_{i+1} -t_{i-1} \}$ 
and $\eps_N \equiv\ sup_{i<N}\{|1 \ -\ (t_{i+1} -t_{i})NF'(t_i)\} $.




We now evaluate the expected number of false $\ell$-change points
for a sequence of kernel 
estimates of $f^{(\ell)}(t)$.
We restrict to independent
$Gaussian$ errors: 
$\eps_i \sim N(0,\sigma^2)$. 
Thus, $\fhlN(t)$ a Gaussian process.
The following assumption rules out nongeneric cases:

\

\noindent
{\bf Assumption B} {\em
Let $f(t) \in C^{\ell+1}[0,1]$ have $K$ $\ell$-change points,
$\{x_1, \ldots x_K\}$, with $f^{(\ell +1)}(x_k)$
$\ne 0 $,
$f^{(\ell )}(0) \ne 0 $ and $f^{(\ell)}(1) \ne 0 $.
Consider a sequence of estimation problems with
independent, normally distributed measurement 
errors, $\eps_i^N$, with variance $\sigma^2$.
Let $\fhlN(t)$ be a sequence of  
kernel estimates of $f^{(\ell)}$, 
on the sequence of intervals, $[\delta_N,1-\delta_N]$.
}

To neglect boundary effects, we take $\delta_N =h_N$ for kernel
estimators and $\delta_N= \delta$ for splines.
For each change point, ${x}_k$, we define 
the change point variance \cite{GM84}:
${\sigma}_{\rm if}^2({x}_k) \equiv\
{ {\bf Var}[\fhlN(x_k)] / |{f^{(\ell+1)}}({x}_k)|^2 }$ . 
The following theorem bounds the probability of a false estimate
of a change point far away from a true change point.

\begin{thm}\label{KZ}
Let 
Assumption B hold and consider a sequence of kernel estimators, 
$\fhlN(t)$, that  satisfy the hypotheses of Lemma \ref{KS}. 
Choose kernel halfwidths, $h_N$, and uncertainty intervals, $w_N$,
such that
$h_N/w_N \rightarrow 0$,  $w_N \rightarrow 0$,
$w_{N,k}^2 N h_N^{2\ell +1} \ge 1$.
The probability, $p_N(w_N)$, that 
$\hat{f}^{(\ell )}_N$ has a false change point outside of a width of 
$w_N$ from the actual $(\ell+1)$-change points satisfies 
\BEQ \label{Kbnd1}
p_N(w_N) \le \
\sum_{k=1}^K {\cal O}
\left( \frac{\sigma_{if}(x_k)}{h_N} 
\exp\left({-w_N^2 \over 2\sigma_{if}^2(x_k)} \right)\ \right) \ ,
\NEQ
where $\sigma_{if}^2(x_k) \rightarrow 
\sigma^2 \|\kappa^{(\ell)}\|^2 \left/ |\flp(x_k)|^2NF'(x_k)h_N^{2\ell +1}
\right.$ 
on the interval $[h_N,1-h_N]$.
\end{thm}

In \cite{Mammen,MMF}, Mammen et al.~derive the number 
of false change points 
for kernel estimation of a probability density.
 We present the analogous
result for regression function estimation. 

\begin{thm}[Analog of \cite{Mammen,MMF}] 
\label{MMFthm}
{Let 
Assumption B hold.
Consider a sequence of kernel smoother estimates $\hat{f}_N$ which 
satisfy the hypotheses of Lemma \ref{KS} with 
$\int_{-1}^1 s \kappa(s)ds$ 
$=0$. 
Let the sequence of kernel halfwidths, $h_N$, satisfy
 $D_N^* N^{1/2} h_N^{\frac{1}{2}} \rightarrow 0 $ and
$0<{\rm liminf}_N h_N N^{\frac{1}{(2\ell+3)}} $
$\le{ \rm limsup}_N h_N N^{\frac{1}{(2\ell+3)}}$ $< \infty $.
The expected number of $\ell$-change points of  $\hat{f}_N$
in the estimation region, $[h_N,1-h_N]$,
is asymptotically
\begin{equation}\label{Mam1} 
{\bf E}[\hat{K}] - K
=  \ 2\sum_{k=1}^K 
H\left(\sqrt{ { | f^{(\ell +1)}(x_k)|^2 NF'(x_k)h^{2\ell+3}
\over \sigma^2\|\kappa^{(\ell +1)}\|^2  } }
\right)  \ + o_{\cal R}(1)\
, 
\end{equation}
where 
$H(z) \equiv \phi(z)/z +\Phi(z) -1$ with $\phi$ and $\Phi$ being the 
Gaussian density. 
If $\flp(t)$ has H\"older smoothness of order $\nu$ for some $0<\nu<1$, 
and $ h_N N^{ {1}/{(2\ell+3)} } \rightarrow 0 $,
then (\ref{Mam1}) remains valid provided that  
$ h_N N^{ {1}/{(2\ell+3+ 2\nu)} } \rightarrow 0 $.
}
\end{thm}

In \cite{Mammen,MMF}, the correction in (\ref{Mam1})
is shown to be $o(1)$  when   
${ \rm limsup}_N h_N N^{\frac{1}{2\ell+3}} < \infty $.
We strengthen this result by showing that (\ref{Mam1}) continues
to represent the leading order asymptotics even when
$ h_N N^{{1}/{(2\ell+3)}} \rightarrow \infty $.

For each change point, ${x}_k$, we define 
the $\alpha$ uncertainty interval by 
$[{x}_k-z_{\alpha}{\sigma}_{\rm if}({x}_k),
{x}_k+z_{\alpha}{\sigma}_{\rm if}({x}_k)]$,
where $z_{\alpha}$ is the two-sided 
$\alpha\left[1 + \ 2 
H( {h^{}_N \|\kappa^{(\ell)}\| \over {\sigma}_{\rm if}({x}_k)    
\ \|\kappa^{(\ell +1)}\|}) \right] $-quantile 
for a normal distribution.
The probability that an empirical change point is more than
$z_{\alpha}{\sigma}_{\rm if}({x}_k)$ away from the 
$k$th actual change point is less than  $\alpha$. 
We consider two change points well resolved if 
the two uncertainty intervals do not overlap.


A similar variance for change point estimation
is given in \cite{Mueller85} where M\"uller shows that
the left-most change point is asymptotically normally distributed
with variance ${\sigma}_{\rm if}^2({x}_k)$. For his result, M\"uller  
imposes stricter requirements on $f^{(\ell +1)}$ and 
 $\hat{f}^{(\ell +1)}$, and does not obtain results pertaining to
the expected number of false change points.


When $f\in C^{m}$, the halfwidth which minimizes the MISE 
scales as $h_N \sim N^{-1/(2m+1)}$. 
Other schemes  for piecewise convex fitting \cite{Li95}
choose the   
kernel halfwidth/smoothing parameter 
to be the smallest value, $h_{cr,K}$, 
that yields only $K$
change points in an unconstrained fit. 
Theorem \ref{MMFthm} 
shows that $h_{cr,K}$ is asymptotically larger than the
halfwidth which minimizes the MISE for $\ell= m, m-1$.
As a result, these  schemes oversmooth.

\section{Data-based Pilot Estimators with Geometric Fidelity} \label{Dadapt}

We consider two-stage estimators that begin by 
estimating $f^{(\ell)}$ and $f^{(\ell+1)}$ using an unconstrained estimate 
with $h_N\gapprox log(N)N^{1/(2\ell +3)}$. 
From the pilot estimate, we 
determine the number, $\Kh$, and approximate locations 
of the change points.
In the second-stage, we perform a
constrained fit; requiring that $\hat{f}^{(\ell)}$ be monotone 
in small regions about each empirical change point.
Since spurious change points asymptotically
occur only in a neighborhood of an actual change point,
the second-stage
fit need only be constrained in  a vanishingly small portion
of the domain asymptotically.

\begin{thm}[Asymptotic MISE for pilot estimation]
\label{DA2thm}
{Let $f(t)$ 
satisfy Assumption B  
and consider a sequence of two-stage estimators.
In the first-stage, let the hypotheses of Theorem \ref{KZ} 
be fulfilled. 
From the first stage estimate, denote the empirical 
$\ell$-change points by $\hat{x}_k, k= 1,\ldots \hat{K}$. 
Choose widths $w_{N,k}$
such that $w_{N,k} \rightarrow 0$, $h_N/w_{N,k} \rightarrow 0$, 
and $w_{N,k}^2 N h_N^{2\ell +1}/\ln(N) \rightarrow \infty$,
where $h_N$ is the first stage halfwidth.
In the second-stage, perform a constrained regression as in (\ref{3.21}),
where the second-stage smoothing parameter, $\lambda_N$, 
satisfies $\lambda_N \rightarrow 0$ and 
$D_N^* \lambda_N^{-5/4m} \rightarrow 0$.  
In the second-stage regression, impose  the constraints that
the second-stage
$\fhlp$ has a single sign 
in the regions $[\hat{x}_k-{w}_{N,k},\hat{x}_k+ w_{N,k}]$
(which matches the sign of $\fhl(\hat{x}_k+ {w_{N,k}})
-\fhl(\hat{x}_k- {w_{N,k}})$ .) 
For $f \in W_{m,2}$,
the second-stage  estimate, $\fh$, satisfies the expected error bounds
of Eq.~(\ref{6.1}) (with different constants).
}
\end{thm}


\noindent
{\bfab Proof.} 
 Theorem \ref{KZ} 
shows that  $\hat{x}_k$
lie within $w_N/2$ of the $x_k$ with probability $1-p_N$, where
$p_N ={\cal O}(\exp(-c_0^2 w_N^2 Nh_N^{2\ell +1}/8\sigma^2))$
and $c_o = \inf_{k}\{|\flp(x_k)|\}$. In the remainder of the
proof, we implicitly neglect this set of measure $p_N$ 
and use  arguments that are valid for large $N$.
By Assumption B, there are no
zeros of $\flp(x_k)$ in $[\hat{x}_k-{w}_{N,k},\hat{x}_k+ w_{N,k}]$
and thus  
$|\fl(\hat{x}_k+ {w_{N,k}})-\fl(\hat{x}_k- {w_{N,k}})| >c_o w_{N,k}$.
Note that $\fhl(\hat{x}_k+ {w_{N,k}})-\fhl(\hat{x}_k- {w_{N,k}})$
has a Gaussian distribution with variance $2 \sigma_N^2(t)$ and a bias
error bounded by ${\cal O}(h_N)$ \cite{GM84}. 
Thus, the sign of $\fhl(\hat{x}_k+ {w_{N,k}})
-\fhl(\hat{x}_k- {w_{N,k}})$ is determined correctly with a probability
of $1-p_N$.
The result follows from  Theorem \ref{MSthm}.
\eopp



The trick of Theorem \ref{DA2thm} 
is to constrain $\fhlp$ 
to be
positive (or negative) in  the uncertainty interval of the estimated 
change points (linear constraints)
rather than constraining $\fhl$ 
to have a single zero around $\hat{x}_j$ (nonlinear constraints). 
Theorem \ref{DA2thm} implies that the second-stage estimate has
no false $\ell$ change points within $\pm w_{N,k}$ of $\xh_k$ with high 
probability. It does not exclude the possibility that false change points
occur outside of   $[\hat{x}_k-{w}_{N,k},\hat{x}_k+ w_{N,k}]$, but
we believe that such false change points seldom occur in practice. 
We believe that it is adequate to eliminate false inflection points
in the regions where they occur in the unconstrained nonparametric
estimates.


Asymptotically, the zeros of $\fhlp$ will occur in clusters with
an odd number of zeros. If a cluster with an even number of zeros
occurs in the first stage, it is spurious (with high asymptotic
probability). We recommend imposing the constraint that the second-stage
$\fhl$ (not $\fhlp$) has a single sign in each neighborhood
where an even number  of change points of the first-stage occur.

For data adaptive methods, we modify Theorem \ref{DA2thm} slightly:

\begin{cor} 
\label{GCVthm0}
The hypotheses of Theorem \ref{DA2thm} on the first-stage estimate
(such as $D_N^*/h_N^{\ell +2}$ $ \rightarrow 0$,
 $w_{N,k}^2 N h_N^{2\ell +1}/\ln(N) \rightarrow \infty$,
$w_{N,k} \rightarrow 0$ and $h_N/w_{N,k} \rightarrow 0$) 
need only be true with probability $(1-p_N)$,
for the conclusion of Theorem \ref{DA2thm} to be valid,
where $p_N$ satisfies $p_N/ \lambda_N \rightarrow 0$
and $p_N N\lambda^{1/ (2m)} 
\rightarrow 0$.
\end{cor}

Let $h_{GCV}$ denote the smoothing bandwidth chosen by generalized 
cross-validation. Under certain conditions  \cite{HHM}, it can be shown that
$h_{GCV}$ has an asymptotically normal distribution with mean
$cN^{-\beta}$, where $c$ and $\beta$
depend on $f$ and  the kernel shape.
For the first-stage halfwidth, we propose using
$h_N =\iota(N) h_{GCV}$, where
$ \iota(N)$ is chosen such that $\iota(N)h_{GCV}$ satisfies
$\ref{GCVthm0}$. If $h_{GCV} \rightarrow cN^{-1/(2\ell+1)}$,
we recommend choosing $ \iota(N) \approx log(N)N^{\alpha}$
with  $\alpha = {1}/{(2\ell +1)} - {1}/{(2\ell+3)}$.
This scaling corresponds to a uniformly consistent estimate
of $\flp$ \cite{GM84}.
The overall moral is:
the smoothing level chosen by GCV is asymptotically optimal for estimating
functions, but derivative estimation requires more smoothing.


Numerical implementations \cite{VW}
of constrained least squares usually apply
the active set method of quadratic programming. 
The constrained smoothing spline regression reduces to a finite 
dimensional minimization when the constraints are on the
$m$th derivative. Since we constrain $\fhlpN(t)$ in 
each neighborhood of an estimated $\ell$-change point,
our algorithm is most readily implemented for $\ell= m-1$
using the duality result of \cite{Riedel1}.

To reduce the number
of constraints, we seek to minimize the length of the constraint intervals,
$w_{N,k}$. When the hypotheses of Theorem \ref{MMFthm} are satisfied
with $h_N N^{\frac{1}{2\ell+3}}\rightarrow \infty $,
we can estimate the uncertainty interval for inflection points
by $ 
\hat{\sigma}^2_{\rm if}(\hat{x}_k)\ \equiv \ 
\sigma^2\|\kappa^{(\ell)} \|^2 /  \left[
|\hat{f}^{(\ell+1)}(\hat{x}_k)|^2 \right.$
$N$ $\left. F'(\hat{x}_k) h_N^{2\ell +1} \right]
 \ .$
We recommend choosing the constraint width such that $w_{N,k} >>
  \hat{\sigma}^2_{\rm if}(\hat{x}_k)$. (For smaller constraint widths,
Theorem \ref{DA2thm} is true, but  false inflection points can still
occur near the actual inflection point.)

\section{Potential Extensions} \label{Pot Ext}

\begin{itemize}
\item
At present, we cannot exclude the posibility of false change points arising
in the second stage estimate away from the constraint intervals. We believe
that this is a technical gap in our analysis and not a practical difficulty.
The risk of false change points can be reduced by choosing larger
intervals to impose the constraints on $\fhlp$. Let $\{\hat{u}_j\}$ denote
the zeros of $\fhlp$ in the  first stage. 
Let $u_j$ and $u_{j+1}$ being the two closest zeros of $\fhlp$ to $x_k$   
with  $u_j < x_k <u_{j+1}$. A judicous choice of constraint intervals is
$[ (x_k - u_j)/2, (x_k + u_{j+1} )/2 ]$, which gives large constraint
intervals with only a small chance of  imposing an incorrect constraint 
on the second-stage $\fhlp$. 

\item
The pilot method suppresses false zeros of  $\fl(t)$, but does
not suppress false zeros of  $\fl(t)-c$ where $c$ is a nonzero real number.
It may be more desirable to apply the pilot estimator to
$\fl(t)- q(t)$, where $q(t)$ is a prescribed function possibly involving
a small number of empirically estimated free parameters. 
The constraints in the second stage will virtually never need to be imposed
if $\fl(t)$ is always positive or negative. Thus,
we suggest centering  $\fl(t)$ about zero
by subtracting off a polynomial fit of order $\ell$
(and thereby centering  $\fl(t)$ about zero) 
prior to applying our two-stage estimator.

\item
Asymptotically, smoothing splines are equivalent to kernel smoothers
\cite{Cox,Silverman84}. Using this convergence,
analogous results to Theorems \ref{KZ}, \ref{MMFthm} and \ref{DA2thm}
can be proved for when smoothing splines are used in the first stage estimate.

\item
It is tempting to try the pilot estimation procedure using  
local polynomial regression (LPR) in the second stage. 
Unfortunately, there is a difficulty
with shape constrained LPR. Let $\hat{f}(x) \sim a_0(t) +
a_1(t)[x-t] + a_2(t)[x-t]^2/2$ for $|x-t| <h$. If $a_2(t)$ is constrained
to be nonnegative, $a_0(t)$ need not be convex because LPR does not
require $a''_0(t) = a_2(t)$.

\item
Our data adaptive convergence results in Sec.~\ref{WLEsect}
and Sec.~\ref{Dadapt} are for quadratic estimation with Gaussian errors.  
The results should be extendable to non-Gaussian errors using the
central limit theorem and Brownian bridges as in \cite{MMF}.
\end{itemize}

\section{Piecewise Convex Information Criterion} \label{PCICSec}

Instead of the two stage pilot estimator, we now propose a second 
class of estimators which penalize both smoothness and the number of
change points.
Information/discrepancy criteria are used to measure whether the improvement 
in the goodness of fit is sufficient to justify using additional 
free parameters. Both the number of free parameters and their values are 
optimized with respect to the discrepancy criterion, $d(\hat{f}, \{y_i\})$.
Let $\hat{\sigma}^2$ be a measure of the average residual error:
$
\hat{\sigma}^2 (\hat{f} ,\{ y_i \} ) = {1\over N\sigma^2 }
\sum_{i=1}^N [y_i -\hat{f} (t_i  )]^2
$,
or its $L_1$ analog. Typical discrepancy functions are
\BEQ \label{RiceEQ}
d^I  (\hat{f} ,\{ y_i \} ) =\hat{\sigma}^2 \left/
[1-(\gamma_1 p /N )]^2 \right. \  \
{\rm and} \ \ 
\NEQ 
\BEQ
d^B  (\hat{f} ,\{ y_i \} ) = \hat{\sigma}^2
[1+ (\gamma_2 p 
\ln (N)/N)] \ \ ,
\NEQ
\ni
where $p$ is the effective number of free parameters 
in the smoothing spline fit. 
For $\gamma_1 =1$, $d^I  (\hat{f} ,\{ y_i \} )$ is 
generalized cross-validation (GCV)
which has the same asymptotic behavior as 
the Akaike information criterion. 
For  $\gamma_2 =1$, $d^B$ is the Bayesian/Schwartz information criterion.
For a nested family of models, $\gamma_2 =1$ is appropriate 
while $\gamma_2 =2$ corresponds to  a nonnested family with
$ 2 {N \choose K}$
candidate models at the $k$th level.
In very specialized settings in regression theory and time series, it has been
shown that functions like $d^I$ are asymptotically
efficient while those like $d^B$ are asymptotically consistent.
In other words, 
using $d^I$-like criteria will
asymptotically minimize the expected error at the cost of not always
yielding the correct model. In contrast, the Bayesian criteria will
asymptotically yield the correct model at the cost of having a larger
expected error.

Our goal is to consistently select the number of convexity change
points and efficiently estimate the model subject to the change point
restrictions. Therefore, we propose the following $new$ discrepancy criterion:
\BEQ \label{PCICeq}
PCIC = \sigma^2 (\hat{f} ,\{ y_i \} ) \left[
{1+\gamma_2 K \ln (N) /N \over \left(1-\gamma_1 p/N \right)^2 } \right] \ ,
\NEQ
\ni
where $K$ is the number of convexity change points and $p$ is the
number of free parameters. 
PCIC stands for Piecewise Convex Information Criterion.
In selecting the positions of the $K$ change points, there are essentially
$2 {N\choose K } $ 
possible combinations of change point locations if we categorize the
change points by the nearest measurement location. 
Thus, our default values are $\gamma_1 =1$ and $\gamma_2 =2$.

We motivate PCIC:
to add a change point requires an improvement in the residual square error
of $O(\sigma^2 \ln (N) )$, which corresponds to an asymptotically
consistent estimate. If the additional knot does not increase the
number of change points, it will be added if the residual error decreases
by $\gamma_1\sigma^2 $. Presently, PCIC is purely a heuristic
principle. We conjecture that it consistently selects the
number of change points and is asymptotically efficient
within the class of methods that are asymptotically consistent
with regards to convexity change points. 

\section{Summary}

Theorem \ref{MMFthm} shows that for $\ell= m$ and $\ell = m-1$,
the amount of smoothing necessary for geometric fidelity is larger
than the optimal value for minimizing the mean integrated square
error. Therefore, we have considered two-stage estimators 
which estimate the $\ell$-change points and their uncertainty
intervals in the first-stage. In the second-stage, a constrained
smoothing spline fit is applied using a data-adaptive estimate
of the smoothing parameter.

Our main result is that such
two-stage schemes achieve the same asymptotic rate of convergence
as standard methods such as GCV that do not guarantee geometric fidelity.
We prove this result incrementally. 
Theorem \ref{MSthm} evaluates an acceptable rate of failure
for imposing the wrong constraints.
Theorem \ref{DA2thm} proves the asymptotic MISE result when the
change points are estimated in the first-stage while
the widths of the constraint intervals and the smoothing parameter
in the first-stage satisfy certain scaling bounds.
The second-stage estimates have no false change points in the
regions where the unconstrained estimators have all of their false change
points with probability approaching unity. 

 Linear constraints are necessary only in small neighborhoods
about each $\ell$-change point asymptotically. 
This suggest that the ratio of the
MISE from our two stage estimate to that of kernel smoothers or spline tends
to one. 
Our estimators should be useful in situations
where obtaining the correct shape is important and computational costs
are not an issue. 
Piecewise convex fitting may offer larger potential gains in the MISE
in small sample situations because the $a priori$ knowledge that
there are only a small number of inflections points should be
of more value when less data is available. Numerical simulations
are underway and will be discussed elsewhere.

\medskip

{\bf Acknowledgments:}
{We thank the referee and editor for their help.
Work funded by U.S. Dept.\ of Energy Grant DE-FG02-86ER53223.}

\appendix

\section{Interpolation Inequalities }
\label{DISCREP}

We measure the distance of an arbitrary set of measurement times
to an equi-spaced set of points in terms of the $discrepancy$
as defined in Sec.~\ref{MISE}. 
The discrepancy is useful because it describe how closely 
a discrete sum over an arbitrarily placed set of points approximates
an integral. In this appendix, we summarize this results and present
a new interpolation identity for discrete sums.
 An useful condition is




\noindent
{\bf Assumption 0} {\em 
Assume that the limiting distribution of the measurement locations, 
$F(t)$, is $C^{1}[0,1]$ and $0<c_F<F'(t)<C_F$. 
}

We denote the set of functions of bounded variation by $TV[0,1]$
and the corresponding norm by $\|\cdot \|_{TV}$.
The discrepancy is useful in evaluating the approximation accuracy
of a discrete sum of arbitrarily placed points to the corresponding 
integral:

\begin{thm} [Generalized Koksma \cite{
RiedLet}] 
\label{Kokcor}
Let $g$ be a bounded function of bounded variation, $\|g\|_{TV}$, on $[0,1]$:
$g \in TV[0,1] \cap  L_{\infty}[0,1]$.
Let the star discrepancy be measured by a distribution, $F(t)$,
which  satisfies Assumption 0.
If the discrete sum weights, $\{w_i, i= 1, \ldots N \}$, 
satisfy $|w_i -1| \le C D_N^*$, then
\BEQ
\left| \int_0^1 g(t)dF(t) - \frac{1}{N}\sum_{i=1}^N g(t_i)w_i 
\right| \le \left[ \|g\|_{TV} + C \|g\|_{\infty} \right] D_N^*   \ .
\NEQ
\end{thm}

In our version of Koksma's Theorem, we have added two new effects:
a nonuniform weighting,  $\{w_i, i= 1, \ldots N \}$, and a nonuniform
distribution of points, $dF$.
The total variation of $g(t(F))$ with respect to $dF$ is equal to
the total variation of $g(t)$ with respect to $dt$. 
Theorem \ref{Kokcor} follows from  Koksma's Theorem
by a change of variables.

In the continuous case, the following Sobolev interpolation result
\cite{Ragozin} is well known:

\begin{lemma} 
\label{SOBint}
There exists constants $c_j$ depending only on $m$ such that for all
$g \in W_{m,2}[0,1]$ and $\theta \in [0,1]$: 
\BEQ\label{Sob}
{\theta^{2j}} \int_0^1 |g^{(j)}(s)|^2 ds \le     
c_j \left[\int_0^1 |g(s)|^2 ds \ + \
{\theta^{2m}} \int_0^1 |g^{(m)}(s)|^2 ds \right] \ .
\NEQ
\end{lemma}

Using Koksma's theorem and Lemma \ref{SOBint}, we can arrive at the following
inequalities: 

\begin{cor}
Let $g$ be in $W_{m,2}[0,1]$ and assume the star discrepancy 
satisfies Assumption 0 with $m <N$.
The following interpolation bounds hold:
\BEQ \label{Kint1}
\frac{1}{N}\sum_{i=1}^N g^2(t_i) \ \le \ 
C_1 \int_0^1 g^2(t)dt
\ + \ c_1 D_N^{*^{m}} \int_0^1 |g^{(m)}(s)|^2 ds  \ , \NEQ
where $C_1 = C_F + c_1 + D_N^*$,
and 
\BEQ \label{Kint2} 
\left[c_F - {c_1 D_N^{*^{\delta}} - D_N^*} \right]
\int_0^1 |g(s)|^2 ds  \le  \frac{1}{N}\sum_{i=1}^N g(t_i)^2  
+\ {c_1} 
D_N^{*^{m(1-2\delta)}} \int_0^1 |g^{(m)}(s)|^2 ds  \ , \NEQ
for all $\delta$ in $(0,1/2)$ such that 
$ c_F > {c_1 D_N^{*^{\delta}} + D_N^*}$. 
%
\end{cor}

{\bfab Proof.} 
For $g \in W_{1,2}$, Koksma's Theorem implies
\vskip-.15in
$$ 
\left| \int_0^1 g^2(t)dF(t) - \frac{1}{N}\sum_{i=1}^N g^2(t_i) 
\right| \ \le\ \|g^2\|_{TV} D_N^* \ \le\
\left( \|g\|_{0,2}^2 +\|g\|_{1,2}^2 \right) D_N^*   \ .
$$
\vskip-.15in
We then apply (\ref{Sob}) to $\|g\|_{1,2}^2$ with $\theta = 
|D_N^*|^{1/2 +\delta}$ for arbitrarily small $\delta$.
\vskip-.075in
$$ 
\left| \int_0^1 g^2(t)dF(t) - \frac{1}{N}\sum_{i=1}^N g(t_i)^2 \right| \le  
c_1 D_N^{*^{\delta}} \left[ \|g\|_{0,2}^2 + D_N^{*^{m(1-2\delta)}}
\|g\|_{m,2}^2 \right]
+ \|g\|_{0,2}^2  D_N^*   \ ,
$$ 
\vskip-.075in
yielding the bound (\ref{Kint2}). 

\vskip-.2in
\section{Proof of Theorem \ref{MSthm}} 
\label{PROOF}

\noindent
{\bfab Proof of Theorem \ref{MSthm}.} 
If the constraints are correct, Theorem \ref{Uterthm}  
yields the asymptotic error bound. 
We need to show that misspecified models do not
contribute significantly to the error.
For any realization of the $\{y_i\}$, we have the bound
\begin{eqnarray}     
\| \fh -f\|_V^2 \ &\le & \lambda (\|f\|^2_m + \|\fh \|^2_m ) + 
\frac{1}{N\sigma^2}\sum_i^N |\fh(t_i)-f(t)-\epsilon_i |^2 +\epsilon_i^2
\nn \\
&\le & \lambda \|f\|^2_m + \frac{1}{N\sigma^2} \sum_i (y_i^2 +\epsilon_i^2)
\ \le\ 
\|f\|_V^2 + \frac{1}{N} \sum_i \epsilon_i^2/\sigma^2 \ .
\end{eqnarray}
This paragraph is devoted to bounding the expectation of 
$\sum_i \epsilon_i^2$ over the worst possible set with probability $p_N$.
Note $\sum_i \epsilon_i^2$ has a $\chi_N^2$ distribution 
with  density  
$p_{\chi_N^2}(w) =  w^{N/2-1} \exp(-w/2)/{2^{N/2}\Gamma(N/2)  }$.
We seek to bound $I_1 \equiv \int ^\infty_{\chi_o^2}w dp_{\chi_N^2}(w)$,
where $\chi_o^2(p_N)$ is defined by $\int^\infty_{\chi_o^2}
dp_{\chi_N^2}(w) =p_N $. 
We claim that $I_1  \le 2N p_N$. We assume that $\chi_o^2(p_N)> 1.5N$.
(If $\chi_o^2(p_N)< 1.5N$,
we split the  integral into $w\le 1.5N$ and $w>1.5N$.) 
We define $\wt = w/N$ and use  Sterling's formula to find 
$p_{\chi_N^2}(N\wt)  
\approx \wt^{N/2-1} \exp(-N(\wt-1)/2)/ \sqrt{4\pi N} $.
Evaluating the integrals by Laplace's method  
under the assumption that $p_N<<1$ and $\chi_o^2>N$ 
yields $p_N \approx 2p_{\chi_N^2}(\chi_o^2)\chi_o^2/(\chi_o^2-N+1/2)$,
and $I_1 \approx 2Np_{\chi_N^2}(\chi_o^2)\chi_o^2/(\chi_o^2-N)$.
For $\chi_o^2> 1.5N$, we have 
$I_1 \approx p_N N (\chi_o^2-N+\ 1/2)/(\chi_o^2-N) << .5 Np_N$.


Taking the expectation conditional on $f \notin V$ yields
${\bf E}_{f \notin V} \| \fh -f\|_V^2 \ \le\ \| f\|_V^2 \ +\ 2$.
Since
$\frac{1}{N}\sum_i |f(t_i)|^2 \rightarrow \int_0^1 f(s)^2 dF(s)$,
we select $N$ large enough that 
$\lambda_N  \|f\|^2_m  > \frac{p_N}{N\sigma^2}\sum_i |f(t_i)|^2$.
We now bound the contribution to
$\Ebf [\| \fh -f\|_m^2 ]$  and $\Ebf [\| \fh -f\|_0 ^2]$  
from $f \notin V$: 
\begin{eqnarray} \label{mbnd}
p_N  {\bf E}_{f \notin V} \| \fh -f\|^2_m \ & \le\ & 
\frac{p_N}{\lambda_N} 
\left(2 \ +\
\lambda_N \| f\|_m^2 \  +\ \frac{1}{N\sigma^2} \sum_i^N f(t_i)^2 \right)
\nn \\
& \le & 
\frac{2\sigma^2}{N\lambda^{2m+1\over 2m}} \  +\
[1+ {\cal O}(p_N)] \  \| f\|_m^2 \ ,
\end{eqnarray}
by assumption on $p_N$. To bound ${\bf E}_{f \notin V} \| \fh -f\|^2_0 $,
we apply  Lemma \ref{SOBint} with $\delta = .05$ in (\ref{Kint2})
and $D_N^*$ large enough that 
$\frac{1}{c_F -c_1 D_N^{*^{\delta}} + D_N^*}$
is bounded by a constant, $\gamma$: 
\begin{eqnarray}\label{zbnd}
p_N  {\bf E}_{f \notin V} \| \fh -f\|^2_0 \ & \le\ & 
\gamma {p_N}{\bf E}_{f \notin V} \left[ 
\frac{1}{N} \sum_{i=1}^N |\fh(t_i)-f(t_i)|^2 
+\ c_1 D_N^{*^{.9m}} \| \fh -f\|^2_m
\right] \nn \\
& \le\ & 
\gamma {p_N}\left( \sigma^2 +  \frac{D_N^{*^{.9m}}}{\lambda}
\right)
{\bf E}_{f \notin V} \left[ 
 \| \fh -f\|^2_V\right]
\nn \\
& \le\ & 
\gamma {p_N}\left( \sigma^2 +  \frac{D_N^{*^{.9m}}}{\lambda} \right)
\left(2 \ +\
\lambda_N \| f\|_m^2 \  +\ \frac{1}{N\sigma^2} \sum_{i=1}^N f(t_i)^2 \right)
\nn \\
& \le &\gamma \left( \sigma^2 +1 \right) 
\left(
\frac{\sigma^2}{N\lambda_N^{1\over 2m}} \  +\
\lambda_N
[1+ {\cal O}(p_N)] \  \| f\|_m^2 \right) \ .
\end{eqnarray}
Equations (\ref{mbnd}) and (\ref{zbnd}) are in the form required by
Theorem 4.4 of \cite{Utreras}.
Using Lemma   \ref{Sob} 
and 
duplicating the proof of Theorem 4.6 of \cite{Utreras} 
yields the result.
\eopp


\vspace{8mm}

\begin{center}
Kurt S. Riedel \\
Courant Institute of Mathematical Sciences \\
New York University \\
New York, New York 10012-1185
\end{center}

\end{document}